\documentstyle[pra,aps,psfig, multicol]{revtex} 
\begin{document} 
 
\newcommand{\ado}{a^\dagger_1} 
\newcommand{\adt}{a^\dagger_2} 
\newcommand{\bdo}{b^\dagger_1} 
\newcommand{\bdt}{b^\dagger_2} 
\newcommand{\bra}[1]{\left<#1\right|} 
\newcommand{\ket}[1]{\left|#1\right>} 
\newcommand{\braket}[2]{\left<#1\right.\left|#2\right>} 
\newcommand{\Tr}[1]{Tr\left( #1 \right)}

\title{On Bell measurements  for teleportation} 
\author{N. L\"{u}tkenhaus, J. Calsamiglia, and  K-A. Suominen} 
\address{Helsinki Institute of Physics, PL 9, FIN-00014 
Helsingin yliopisto, Finland} 
\date{\today} 
\draft 
\maketitle 
\begin{abstract} 
In this paper we investigate the possibility to make complete Bell
measurements on a product Hilbert space of two two-level bosonic
systems. We restrict our tools to linear elements, like beam splitters
and phase shifters, delay lines and electronically switched linear
elements, photo-detectors, and  auxiliary bosons. As a result we show that with these tools a never failing  Bell measurement is impossible. 
\end{abstract} 
 
\begin{multicols}{2}[] 
\narrowtext 
 
\section{Introduction} 
Bell measurements project states  of two two-level systems onto the
complete  set   of   orthogonal maximally    entangled  states   (Bell
states). The motivation  to deal with  Bell states comes from the fact
that  they  are key ingredients in  quantum   information. Bell states
provide   quantum correlations which  can be  used in certain striking
applications    such as: teleportation   in  which a  quantum state is
transferred  from one  particle to  another in   a ``disembodied'' way
\cite{bennett93a}, quantum dense coding in  which two bits of information
can  be  communicated by   only   encoding  a single  two-level   system
\cite{bennett92}, and
entanglement   swapping \cite{jian98,zukowski93} which allows  to
 entangle  two  particles that do not have any   common past, 
and  opens a  source full  of new applications since it provides a
  simple way of creating multiparticle entanglement \cite{zeilinger97,bose98}.
     But to take full advantage  of
these applications  one needs to be able  to  prepare and measure Bell
states. The problem of creating Bell states has been solved in optical
implementations  by   using  parametric downconversion  in a  non-linear
crystal \cite{kwiat95}. Particular Bell states   can be prepared  from
any     maximally   entangled     pair by     simple   local   unitary
transformations. The question arises whether it is possible to perform
a complete Bell   measurement with linear devices  (like beam-splitters
and phase shifters). It  is clear that  this can be achieved  once one
has the ability to  perform a controlled NOT  operation (CNOT) on  the
two   systems,  which transforms the   four    Bell states  into  four
disentangled  basis states. In principle we need  to do less. As we
are  not interested in the state  of the system after the measurement,
it can be vandalized by the measurement. The only important thing is
the measurement result identifying unambiguously  a Bell state.

In  an earlier  papers Cerf, Adami,  and Kwiat  \cite{cerf98a} have shown
that it  is possible  to  implement quantum   logic  in purely  linear
optical  systems.  These  operations, however, do   not   operate on a
product of Hilbert spaces of  two systems, instead they operate on
product of Hilbert spaces of two degrees of freedom (polarization and
momentum) of the same system. Therefore these results can
be used to implement  quantum logic circuits but not to
 perform most of the applications mentioned above. 
 For example, in the case of 
 teleportation there have   been two  recent   experimental  realizations
\cite{boschi,bouwmeester97a}.
 Boschi et al. presented results  in which Bell measurement is
realized with  100 \% efficiency using linear  optical  gates, but the
teleported state  has  to  be  prepared  beforehand over one   of  the
entangled photons \cite{boschi} . So,   in some sense  that  scheme 
differs from  the
``genuine''   teleportation since it  does not   have some very crucial
properties, like the ability  to  teleport entangled states  or  mixed
states.  This obstacle  could, of course,  be overcome if one  had the
possibility to  swap the unknown  state  to the EPR photon.  But, this
again requires quantum-quantum interaction (not linear operator).  On
the other hand the Innsbruck experiment can be considered as a ``genuine''
teleportation but it has the important  drawback that it only succeeds
in 50 \% of the  cases (in the remaining  cases the original state  is
destroyed). For  the same reason  the Innsbruck  dense coding experiment
\cite{mattle96} can only reach a   communication rate of $1.58$ bits  per
photon  instead of $2$ bits per photon.     

Recently,  Kwiat  and Weinfurter \cite{kwiat98} have presented a
method  which allows complete Bell   measurements and that operates on
the  product  Hilbert  spaces of two   systems,  but  it   adds a very
restrictive  requirement too. That is,  the particles need to be
entangled in some other degree of freedom  beforehand (so, half of the
job is already done). Notwithstanding, this method still represents an
important  progress since it  allows,   in principle,  to realize  all
applications which fulfill the condition that  the Bell measurement is
performed  over photons which have  already quantum correlations (like
in the case of  quantum dense coding). 

 At this stage  we choose to call
a physical scheme  a Bell analyzer only if  it operates on  product
Hilbert spaces of two two-level systems. 
A generalization to systems with other structure than a two-level system
is the measurement  used in the  teleportation of continuous variables
\cite{braunstein98a} which successfully projects on singlet states.

In this paper we prove that all these  turnabouts are more than 
 justified since we present a no-go-theorem  for Bell analyzer
 for experimentally  accessible measurements involving only linear 
quantum  elements. We now  lay
out the framework for this theorem in a language which clearly has the
experimental situation  of  the teleportation experiment performed  in
Innsbruck  in  mind. This  means  especially  that we  concentrate  on
bosonic input states. Results concerning fermionic  input or input of
distinguishable particles  can  be  found  in   the work   of  Vaidman
\cite{vaidman98a}.   

  The Hilbert space    of the input  states  is
spanned by states describing  two photons coming into  the measurement
from two different spatial directions,  each carrying two polarization
modes. Therefore we  can describe the input  states in a  sub-space of
the excitations of four  modes  with photon creation  operators $\ado,
\adt, \bdo,  \bdt$. Here $a$  and  $b$ refers to  spatial modes, while
``1'' and   ``2'' refer to  polarization  modes. The  Hilbert space of
interest is spanned by the orthonormal set of Bell states given by,  
\begin{eqnarray} 
\ket{\Psi_1}  & =  &\frac{1}{\sqrt{2}}\left( \ado  \bdt   - \adt  \bdo
\right) \ket{0 } \\  
\ket{\Psi_2} &  = &\frac{1}{\sqrt{2}}\left( \ado  \bdt   + \adt \bdo
\right )
 \ket{0} \\  
\ket{\Psi_3} & =
&\frac{1}{\sqrt{2}}\left( \ado \bdo  - \adt \bdt \right)  \ket{0} \\ 
\ket{\Psi_4}   & = &\frac{1}{\sqrt{2}}\left( \ado   \bdo  + \adt  \bdt
\right) \ket{0}   
\end{eqnarray} 
where $\ket{0}$ describes the vacuum state. Although we used spatial
modes and polarization to motivate this form of Bell states, it should
be noted that any two pairs of bosonic creation operators (all four commuting)
can be chosen for the theorem to be valid.  This includes all possible
degrees of freedom of the boson. In the photon case it includes
especially polarization, time, spatial mode and frequency.  For
example, all wave packets containing one photon can be modeled.    
The Bell measurement we are looking for is described by a positive
operator valued measure (POVM) \cite{peres93} given by a collection of
positive operators $F_k$ with $\sum_k F_k = \openone$. Each operator $F_k$
corresponds to one classically distinguishable measurement outcome,
for example that detectors ``1'' and ``2'' out of four detectors go
``click'' and the rest do not. The probability $p_k$ for the outcome
$k$ to occur while the input is being described by density matrix
$\rho$,  is given by $p_k = \Tr{\rho F_k}$.  A Bell measurement with
100 \% efficiency is characterized by the property that all $F_k$ are 
triggered with probability
$\Tr{\rho_{\Psi_i} F_k} \neq 0$ for only one of the four Bell state
inputs $ \rho_{\Psi_i}$ ($i=1,\dots,4$). This allows us to rephrase
the problem as one of {\em distinguishing} between four orthogonal
equally probable Bell states with 100 \% efficiency.

\begin{figure}[] 
\centerline{\psfig{figure=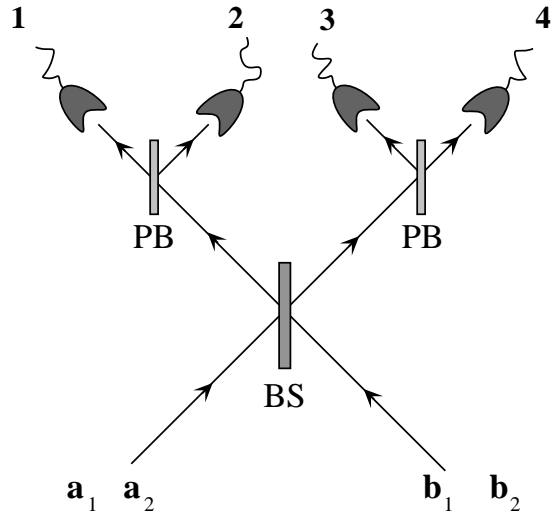,width=3in}}  
\label{innsbruck}
\vspace{0.2cm}
\caption{ The  Innsbruck  detection scheme  uses an  initial  $50/50$
beam-splitter (BS) mixing  modes $\ado$ with  $\bdo$ and  $\adt$  with
$\bdt$. Then  each of the resulting outputs  is separated  from each 
other using a polarizing beam-splitter (PB). }  
\end{figure}

 To illustrate the formalism we  look at the Innsbruck detection
scheme \cite{weinfurter94} (fig. \ref{innsbruck}), which consists of 8
POVM elements, corresponding to the events 
\[ 
\begin{array}{c|c} 
\mbox{detectors going ``click''} & \mbox{could have been} \\ 
 & \mbox{triggered by } 
\\ \hline 
\mbox{``1'' and ``4''} & \Psi_1 \\ 
\mbox{``2'' and ``3''} & \Psi_1 \\ 
\mbox{``1'' and ``2''} & \Psi_2 \\ 
\mbox{``3'' and ``4''} & \Psi_2 \\ 
\mbox{``1'' sees 2 photons} & \Psi_3 \; \mbox{ or }\; \Psi_4\\ 
\mbox{``2'' sees 2 photons} & \Psi_3 \;\mbox{ or }\; \Psi_4\\ 
\mbox{``3'' sees 2 photons} & \Psi_3 \; \mbox{ or }\; \Psi_4\\ 
\mbox{``4'' sees 2 photons} & \Psi_3 \; \mbox{ or }\; \Psi_4 
\end{array} 
\] 

   Only the  first   four events allow assigning  unambiguously  Bell
states   to the  outcomes. The total   fraction  of  these  events for
teleportation,  where all Bell states  are   equally probable, is  $50
\%$.  The state demolishing projection on entangled  states is
indeed possible  using only linear  elements, but  not 100 \%
efficient.  

\section{Description of the considered measurements} 
\begin{figure}
\centerline{\psfig{figure=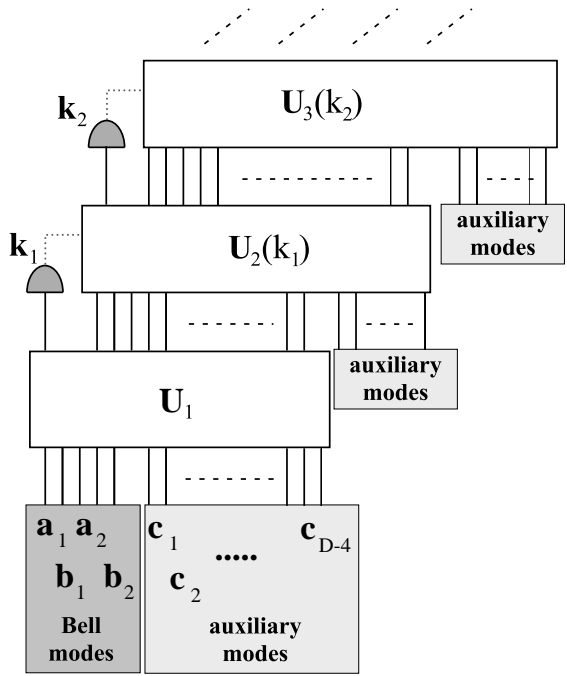,width=3in}}  
\vspace{0.2cm}
\caption{ The general scheme mixes the modes  of the  Bell state with
auxiliary modes (not necessarily   in  the vacuum  state).  Then  one
selected mode is measured and, depending on the  measurement outcome,
the other output modes are mixed  with new modes  and inputs linearly
and  again a mode is selected  to be  measured.  This  process can be
repeated over and over again.}  
\label{generalscheme} 
\end{figure}

Before  we continue we shall describe   our tools more precisely. We
restrict our measurement  apparatus to  linear elements only. This
means that  the vector  of creation operators  of the  input modes is
mapped  by a unitary  matrix onto the vector  of creation operators of
the output modes. Reck et al. \cite{reck94a}  have shown that all these
unitary mappings can be realized  using only beam splitters and  phase
shifters. The  number of modes is not necessarily four: we can couple
to more  modes using  beam-splitters  so that   the  input states  are
described by   the   direct product   of  the  Hilbert  space   of the
Bell states and  the initial state  of  the additional modes.     All
those   modes  are mapped   into  output  modes, where  place
detectors. We  assume these detectors  to be ideal,  so that  they are
described as performing a POVM measurement on the monitored mode where
each  POVM  element $F_k^{(detector)}     = \ket{k}\bra{k}$  is    the
projection onto  a Fock state of that  mode. For experimental reasons,
one would like to  reduce  this to a simpler   detector that can   not
distinguish the number of photons by which it is triggered. The simple
``click''  or ``no click''  detector is  described by  a POVM with two
elements,         $\ket{0}\bra{0}$      and         $\sum_{k=1}^\infty
\ket{k}\bra{k}$. However,  we will  show  that  even a more  fancier
detector does not allow us to implement  a Bell measurement that never
fails.  The  last  tool introduced  here is   the  ability to perform
conditional  measurements. With  that  we  mean  that we monitor   one
selected mode while keeping the other modes in a waiting loop. Then we
can perform some linear operation on  the remaining modes depending on
the outcome of the measurement  with all the tools described  above. 
The  general strategy  is  shown schematically  in
figure \ref{generalscheme}.    

 Vaidman and Yoran \cite{vaidman98a} have arrived to the
conclusion that a  Bell state analyzer can not be build using only
linear devices,  but their  measurement apparatus does only a
very restrictive type of measurement.
It  is not allowed to make use of auxiliary photons
 and no conditional measurements are allowed neither.
Both tools might be very useful and we do not see any
essential reason to disregard them.
 For instance  the apparatus proposed by Vaidman and Yoran can not 
distinguish between the four disentangled basis-states of the from,
\[ 
 \ket{\uparrow} \ket{\leftarrow}, \ket{\uparrow} \ket{\rightarrow}, 
  \ket{\downarrow} \ket{\uparrow}, \ket{\downarrow} \ket{\downarrow}
\]
for which a conditional measurement is needed.

\section{Criticism to   a priori arguments against linear Bell measurements} 

 Intuitively, one needs to  operate a ``non-linear''  measuring device
 to perform  Bell measurements in the  sense that one two-level has to
 interact with the other. 
In the case of photons there is no direct interaction between
 them. One can try to couple them through a third system such as an
 atom \cite{Torm96} or map the state of the photons into atom or ion states
 and perform there the desired measurement \cite{vEnk97}.
These schemes are closest  to the simple idea of performing
 a CNOT  operation, a Hadamard  transform and than projecting on the
 disentangled base, but they bring up a whole new range of problems
 (e.g. weak coupling, decoherence, pulse shape design) 
that breaks with  the idea of having simple and controlled
 ``table-top''  optical implementations of Quantum Information applications.
Therefore it is worth checking the possibility of performing
it  by linear means.

It is true that linear operations can not make the two input photons
interact, they can only make them interfere. Therefore the unitary
transformation $U_L$ is separable in the sense that it can be written in
 terms of a
unitary operation $U$ over each photon, and of
course a CNOT can not be performed by these means ($U_L=U\otimes U$ acts on the
symmetric subspace of the  single photon Hilbert space product
${\cal H}_1\otimes {\cal H}_1$,  dim$(U)>2$ ). 
Even if this kind  of operation preserves the entanglement,
 the Hilbert space might be large enough to span outputs which trigger
 different combination of detectors for different input  Bell state. 
 
\section{No-Go Theorem} 
We now show that it is not possible to construct  a Bell measurement
 using only  the tools mentioned above to realize a measurement,
 for which all POVM elements are  projections  on one of the four 
orthogonal Bell states. 
 
\begin{figure}
\centerline{\psfig{figure=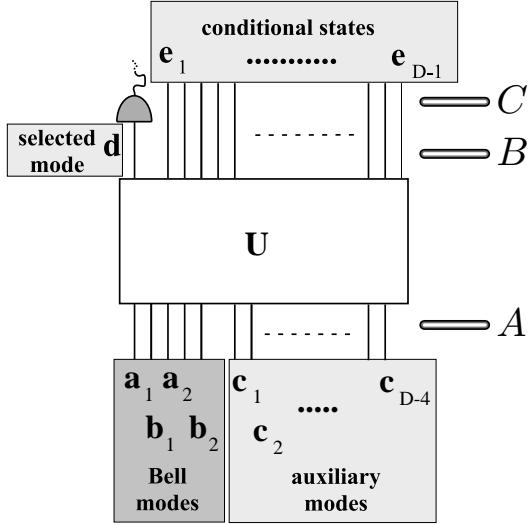,width=3in}} 
\vspace{0.2cm}
\caption{ The initial step takes the input state at 
stage A from the input mode description via the linear transformation 
$U$ to the output mode description at stage B. Depending on the 
detected photon number in mode $d$ we find different conditional state 
for the four Bell state inputs at stage C.} 
\label{specialscheme} 
\end{figure} 

To do so we concentrate  on the first step of our measurement set-up: 
We measure the  photon number in one selected mode $d$ (see figure
\ref{specialscheme}).
 For each result we will find the remaining  modes in four
conditional states corresponding to each Bell state input.
 We then show that there is always at least one photon number
detection event in the first mode that leads to  non-orthogonal 
(i.e. not distinguishable) conditional states in the remaining modes.  
 
In stage A (fig. \ref{specialscheme}) the input state can be described as a product of two polynomials in the creation operators of the auxiliary and the Bell states modes respectively acting onto the vacuum (denoted by $\ket{0}$): 
\[ 
\ket{\Psi^{(total)}_i} = P_{aux}\left(c^\dagger_j\right) P_{\Psi_i}\left(\ado,\adt,\bdo,\bdt\right)\ket{0}  
\] 

Since we use detectors with photon number resolution it is enough to assume that the auxiliary input is in a state of definite photon number. Then $P_{aux}\left(c^\dagger_j\right)$  contains only products of a fixed number of creation operators, and $P_{Bell}\left(\ado,\adt,\bdo,\bdt\right)$ contains only products of two creation operators. 
Now  the modes of the Bell state
input $a_1,a_2,b_1,b_2$ and the auxiliary modes $c_j$
 are linearly mapped by the unitary transformation $U$ into the output modes
$d$ and $e_k$. At stage B the state is described by 
\[ 
\ket{\Psi_i^{(total)}} = \tilde{P}_{aux}\left(d^\dagger,e^\dagger_k\right)  
\tilde{P}_{\Psi_i}\left(d^\dagger,e^\dagger_k\right)\ket{0}  
\] 
We expand the two polynomials in powers of $d^\dagger$ as 
\begin{eqnarray} 
\tilde{P}_{aux}\left(d^\dagger,e^\dagger_k\right) & = & \left( 
d^\dagger\right)^{N_{aux}} 
\tilde{Q}_{aux}\left(e^\dagger_k\right)+ \dots \\ 
\tilde{P}_{\Psi_i}\left(d^\dagger,e^\dagger_k\right) & = & \left( 
d^\dagger\right)^{N_{Bell}} 
\tilde{Q}_{\Psi_i}\left(e^\dagger_k\right)+ \dots 
\end{eqnarray} 
$N_{Bell}$ is defined as the maximal order in $d^\dagger$ among the
four polynomials $\tilde{P}_{\Psi_i}$ and it is 
independent of the index $i$. As a consequence, the polynomials 
$\tilde{Q}_{\Psi_i}$ can be zero for some $i$. Similarly $N_{aux}$
is defined as the order in  $d^\dagger$ of the polynomia
$\tilde{P}_{aux}$.

In the mode $d$ we will  find a range of photon numbers. To
prove the theorem it suffices to see that for any of these events the
conditional states $\ket{\Phi_i^{(total)}}$  that arise for each of the
Bell states, are not perfectly distinguishable. 
We concentrate on the measurement outcomes in this mode which leads
 to the maximum photon number detected in that mode, $N= N_{aux} +
N_{Bell}$.
 The states $\ket{\Phi_i^{(total)}}$ of the remaining modes
conditioned on the occurrence of this event  is then given by 
\begin{equation} 
\label{condstate} 
\ket{\Phi_i^{(total)}} = 
\tilde{Q}_{aux}\left(e^\dagger_j\right) 
\tilde{Q}_{\Psi_i}\left(e^\dagger_j\right)\ket{0} \; . 
\end{equation} 
 
The reason of starting out from the event of detecting the $N$ photons in the selected mode $d$, is that the  
problem reduces to a much simpler form in which the measuring apparatus is not allowed to make use of auxiliary photons. 
That is, by imposing the orthogonality condition of the conditional states on this particular event, we prove  
that the contribution $\tilde{Q}_{aux}\left(e^\dagger_j\right)$ 
of the auxiliary photons can not make non-orthogonal states orthogonal in the sense that two conditional states $\ket{\Phi_i^{(total)}}$ are orthogonal if and only if the  the states,\[ 
\ket{\Phi_i} = 
\tilde{Q}_{\Psi_i}\left(d^\dagger,e^\dagger_j\right)\ket{0} 
\] 
are orthogonal.  
 
To prove this statement we observe that the overlap of two 
conditional states belonging to different Bell state input $i$ and $j$ 
is given by 
\begin{eqnarray} 
\braket{\Phi_i^{(total)}}{\Phi_j^{(total)}} 
& = & 
\bra{0}\tilde{Q}_{aux}^\dagger 
\tilde{Q}_{\Psi_i}^\dagger 
 \tilde{Q}_{aux} \tilde{Q}_{\Psi_j} 
\ket{0}  \nonumber \\ 
& = & \sum_{\overline{n}} \bra{0}\tilde{Q}_{aux}^\dagger 
\tilde{Q}_{aux}\ket{\overline{n}}\bra{\overline{n}} 
 {\tilde{Q}_{\Psi_i}}^\dagger \tilde{Q}_{\Psi_j} 
\ket{0} \nonumber \\ 
& = &  \bra{0}{\tilde{Q}_{aux}}^\dagger 
\tilde{Q}_{aux}\ket{0}\bra{0} 
 {\tilde{Q}_{\Psi_i}}^\dagger \tilde{Q}_{\Psi_j} 
\ket{0} . 
\end{eqnarray} 
The first step makes use of the commutativity of
$\tilde{Q}_{\Psi_i}^\dagger$ and $\tilde{Q}_{aux}$ following the
commutativity of the two set of creation operators for the auxiliary
modes and the Bell modes. Furthermore, the first step   inserts the
identity operator of the Fock space for all involved modes. We denote
by $\overline{n}$ the vector of photon numbers in each involved mode.
The second step then uses the fact that only one of these terms is nonzero.
 This is  a consequence of $\tilde{Q}_{\Psi_j}\ket{0}$ being a
state with total photon number $2$ while the conjugate state 
$\bra{\overline{n}} \tilde{Q}_{\Psi_i}^\dagger$ is a two photon state if 
and only if $\bra{\overline{n}} = \bra{0}$. 
 
Now that is clear that the use of auxiliary photons does not 
provide any help in building a  Bell state analyzer, it is much easier
to check if the orthogonality condition of the conditional states
 is fulfilled when only one or two photons are detected in the
selected mode $d$. To do this, we introduce a  formalism for the
  linear mapping of modes. 
 
Consider the unnormalized input state 
\begin{eqnarray} 
\ket{\Psi} &=&  
\frac{\mu_1}{\sqrt{2}}\left(\ado \bdo + \adt \bdt \right)+ 
\frac{\mu_2}{\sqrt{2}}\left( \ado \bdo + \adt \bdt \right) \\ 
& &+ \frac{\mu_3}{\sqrt{2}}\left( \ado \bdt - \adt \bdo \right)+ 
\frac{\mu_4}{\sqrt{2}}\left( \ado \bdt - \adt \bdo  \right) 
\ket{0} \; . \nonumber 
\end{eqnarray} 
By choosing one of the weights $\mu_i$ as one and the others as zero,
we recover the four Bell states. This state can be written with the
help of a symmetric real matrix ${\bf M}$ as 
\[ 
\ket{\Psi} = (\ado, \adt, \bdo, \bdt,...) {\bf M} (\ado, \adt, \bdo,
\bdt,...)^T \ket{0}  
\] 
with  
\[ 
{\bf M} = 2^{\frac{3}{2}}\left( 
\begin{array}{ccccccc} 
0 & 0 & \mu_1 + \mu_2 & \mu_3 + \mu_4 & 0 &...&0 \\ 
0 & 0 & \mu_3 - \mu_4 & \mu_1 - \mu_2& 0 &...&0  \\ 
\mu_1 + \mu_2 &\mu_3 - \mu_4 & 0 & 0 & 0 &...&0 \\ 
\mu_3 + \mu_4 & \mu_1 - \mu_2 &  0 & 0& 0 &...&0 \\ 
0 & 0& 0 & 0 & 0 & ...& 0 \\ 
: & : & : & : & :& : & : 
\end{array} 
\right). 
\] 
A linear transformation of the modes is now equivalent to the transformation 
\[ 
{\bf \tilde{M}} = U^T {\bf M} U 
\] 
for some matrix $U$ of dimension $ D \times D$ (with $ D \geq 4$) 
satisfying  $U U^\dagger = \openone$. The choice $D \ge 4$ 
corresponds to an enlargement of the number of modes due to additional
unexcited input modes of beam-splitters.
 The output modes are now $d, e_1, ..., e_{D-1}$.
 The entries of the matrix $\tilde{M}$ reveal the distinguishability
 of the Bell states in the following way: 
if two photons are detected in the mode $d$ then the presence
 of $\mu_i$ in the matrix element 
$\tilde{M}_{11}$ reveals which Bell states $\Psi_i$ could have contributed to this event. For all Bell states that contribute, the conditional state of the remaining modes is vacuum. It turns out that  this event can not be attributed to a single Bell state. 
To prove this statement we calculate $\tilde{M}_{11}$ with a general first column of the matrix $U$ given as ${\bf v_1} = (a , b , c, d,...)^T$: 
\begin{eqnarray} 
\tilde{M}_{11}& = &{\bf v_1}^T {\bf M}  {\bf v_1} \\ 
 & = & \frac{1}{\sqrt{2}} \mu_1 (a  \; c  + b  \;d) + \frac{1}{\sqrt{2}} \mu_2 (a \;c  -b  \;d) \nonumber \\ 
 & &  + \frac{1}{\sqrt{2}} \mu_3 (a  \; d + b  \; c) + \frac{1}{\sqrt{2}} \mu_4 (a \; d - b\;c). \nonumber  
\end{eqnarray} 
To be able to attribute the event of two photons in one mode 
unambiguously  to one Bell state, one and only one of the coefficients of the $\mu_i$'s should be non zero. 
It is easily verified that this condition can not be satisfied.  
 
If we impose that three of the coefficients vanish we obtain two
 possible solutions, 
\begin{eqnarray} 
a = 0 &,& b = 0  \;  \; \;    \forall c, d \; \; \mbox{ i.e. } {\bf v_1} =(0,0, c, d)\\ 
c = 0 &,& d = 0  \;  \; \;  \forall a, b \; \; \mbox{ i.e. } {\bf v_1} =(a, b,0,0). \nonumber  
\end{eqnarray} 
But for both solutions $\tilde{M}_{11}=0$. Therefore a perfect Bell
 analyzer can never detect two photons in the selected mode.
 Now we have left only
 the case where only one photon is detected. 
 
After a single photon detection at mode $d$, the first line of
${\bf \tilde{M}}$, denoted by ${\bf \tilde{M}_{1,i}}$ tells us the
state of the remaining modes. Their state is derived from the unnormalized state 
\[ 
\ket{\Phi} =  {\bf \tilde{M}_{1,i}} (d^\dagger,e_1^\dagger, ..., e_{D-1}^\dagger)^T  
\] 
by choosing, as before, one of the $\mu_i$ to one, and the rest to 
zero. We have shown above that the first column of $U$ is of the form ${\bf v_1} =(a, b,0,0)$ or ${\bf v_1} =(0,0, c, d)$ in order to avoid two photons 
entering  the selected mode. Due to the symmetry of the 
problem we can restrict ourselves to the first situation, ${\bf v_1} =(a, b,0,0)$. We now write $U$ in the form 
\[ 
U = \left(  
\begin{array}{cc} 
a & {\bf a_R} \\ 
b& {\bf b_R} \\ 
0 & {\bf c_R} \\ 
0 & {\bf d_R} \\ 
: & : \\ 
\end{array} 
\right) 
\] 
Here  ${\bf a_R}, {\bf b_R}, {\bf c_R}, {\bf d_R}$ are $D-1$ dimensional row vectors. Then  
${\bf \tilde{M}_{1,i}}$ is given by 
\begin{eqnarray} 
{\bf \tilde{M}_{1,i}}& =&\frac{1}{2 \sqrt{2}} \left(0, \mu_1 (a{\bf c_R} + b {\bf d_R}) + 
 \mu_2 (a{\bf c_R} - b {\bf d_R}) + \right.\nonumber \\ 
 & & \left.  \mu_3 (b {\bf c_R} + a{\bf d_R}) + \mu_4 (b {\bf c_R} - a{\bf d_R}) \right) 
\end{eqnarray} 
From this it follows that the conditional states are (up to 
normalization) 
\begin{eqnarray} 
\ket{\Psi_1} & = &(a\; {\bf c_R} + b \; {\bf d_R}) {\bf e^\dagger} \ket{0} \\ 
\ket{\Psi_2} & = &(a\; {\bf c_R} - b \; {\bf d_R}) {\bf e^\dagger} \ket{0} \\ 
\ket{\Psi_3} & = &(a\; {\bf d_R} + b \; {\bf c_R}) {\bf e^\dagger} \ket{0} \\ 
\ket{\Psi_4} & = &(a\; {\bf d_R} - b \; {\bf c_R}) {\bf e^\dagger} \ket{0}  
\end{eqnarray} 
with the vector of creation operators ${\bf e^\dagger}=(e_1^\dagger, 
\dots, e_{D-1}^\dagger)^T$. 
The six different overlaps between these states are (up to the missing normalization factors): 
\begin{eqnarray} 
\braket{{\Psi_1}}{{\Psi_2}} & = & |a|^2 |{\bf c_R}|^2 - |b|^2 |{\bf 
d_R}|^2 \\ 
\braket{{\Psi_1}}{{\Psi_3}} & = & a^\ast b  |{\bf c_R}|^2 + 
b^\ast a|{\bf d_R}|^2 \\ 
\braket{{\Psi_1}}{{\Psi_4}} & = & b^\ast a|{\bf d_R}|^2 -
a^\ast b  |{\bf c_R}|^2  \\ 
\braket{{\Psi_2}}{{\Psi_3}} & = & a^\ast b  |{\bf c_R}|^2 - 
b^\ast a|{\bf d_R}|^2 \\ 
\braket{{\Psi_2}}{{\Psi_4}} & = &- a^\ast b  |{\bf c_R}|^2 - 
b^\ast a|{\bf d_R}|^2 \\ 
\braket{{\Psi_3}}{{\Psi_4}} & = &|a|^2 |{\bf d_R}|^2 - 
|b|^2 |{\bf c_R}|^2  \; . 
\end{eqnarray} 
These overlaps are zero if, 
\begin{eqnarray} 
(|a|^2- |b|^2) (|{\bf c_R}|^2 + |{\bf d_R}|^2) & = & 0 \\ 
(|a|^2+ |b|^2) (|{\bf c_R}|^2 - |{\bf d_R}|^2) & = & 0 \\ 
a^\ast b  \; |{\bf c_R}|^2 & = & 0 \\ 
b^\ast a \; |{\bf d_R}|^2 & = & 0 \; . 
\end{eqnarray} 
Since the column vector ${\bf v_1}$ can not be a zero vector ($|a|^2+ |b|^2 \neq 0$) this simplifies to 
\begin{eqnarray} 
 |{\bf c_R}|^2 & = & |{\bf d_R}|^2 \\
2(|a|^2- |b|^2) |{\bf c_R}|^2  & = & 0 \\ 
b^\ast a|{\bf c_R}|^2 & = & 0 \; . 
\end{eqnarray} 
from which we can conclude that $|{\bf c_R}|^2 = |{\bf d_R}|^2= 0$.
 But for this 
choice the matrix $U$ does not have rank $4$ and so the restriction on 
$U$ given by $U U^\dagger = \openone$ can no longer be 
satisfied. Obviously now we can discard the only remaining case;  the zero
 photon case  represents a bad choice of the mode $d$
 since it would be disconnected from the incoming Bell modes.
This is the final blow to the attempt to do Bell 
measurements  with linear elements.  
 
\section{Conclusion} 
In this paper we have shown that no experimental set-up using only
linear elements can implement a Bell state analyzer. Even the ``non-linear
experimentalist'' performing photon number measurements and acting
conditioned on the measurement result can not achieve a Bell
measurement which never fails.  Included in the proof is the
possibility to insert entangled states in auxiliary modes into the measurement device.  

Recently there has been another proof of this no-go theorem
\cite{vaidman98a} and some proposals to surmount the theorem 
\cite{boschi,kwiat98,braunstein98a,weinfurter94}.
In this paper we have discussed their oversights or drawbacks
 and explained why the theorem does not apply to them.

 The remaining open question is the one for the maximal fraction of 
successful Bell measurements. The Innsbruck scheme gives $50 \%$. 
It should be noted, that in principle all numbers between $50 \%$ and,
in a limit, $100 \%$ can be allowed by a POVM measurement, which
either gives the correct Bell state or gives an inconclusive result.  
Something that can help to gain some insight on the problem 
is to investigate the possibility of projecting with
(or  asymptotically close to) 100\% 
efficiency over a not maximally entangled base (but still with some
 entanglement).
 
The fact that the first step in our proof was to rule out the use of an
 auxiliary system, does not mean that it could not be a very useful tool
 when considering the case of obtaining an efficiency bigger than 50
 \%. Following the same procedure than in this proof,
 and trying to  evaluate
 the maximum distinguishability of the conditional states
 \cite{chefles98}
 that appear in each stage, could be a way to obtain
 the real upper-bound to the Bell measurement efficiency. 

\section{Acknowledgments}
The authors thank the organizers of the ISI (Italy)
and Benasque Center for Physics (Spain) workshops on quantum
computation and quantum information held in summer 1998
which brought us in contact with the works by Vaidman and Yoran
 \cite{vaidman98a}, and Kwiat and Weinfurter \cite{kwiat98}.
We also thank L.\ Vaidman and M.\ Plenio for useful discussions and the
 Academy of Finland for financial support.

\end{multicols}

\end{document}